\documentclass[10pt,pre]{article}\oddsidemargin-1.cm \textwidth 18cm\headheight-2.5cm \textheight 24.5cm\newcounter{benji}
\usepackage[round]{natbib} 
\usepackage{psfig,epsfig}
\newcommand{\be}{\begin{equation}}
\newcommand{\ee}{\end{equation}}
\newcommand{\ba}{\begin{eqnarray}}
\newcommand{\ea}{\end{eqnarray}}
\newcommand{\rd}{R\&D }
\newcommand{\intl}{\int\limits}
\newcommand{\liml}{\lim\limits}

\newcommand{\lr}[1]{\langle #1 \rangle}
\newcommand{\LR}[1]{\Big\langle #1 \Big\rangle}
\newcommand{\te}{\textstyle}
\newcommand{\nz}{\nonumber}
\newcommand{\bi}[1]{Fig.~\ref{fig:#1}}
\newcommand{\e}[1]{eq.~(\ref{#1})}

\begin{document}
\ifcase\value{benji} \setlength{\baselineskip}{1.8em} \or \setlength{\baselineskip}{1.5em} \fi
\vskip3em

\begin{center}
{\Large Comment on:
{\sl Characterization of subthreshold voltage fluctuations in  neuronal membranes} 
 by M.~Rudolph and A.~Destexhe}\\[3em]

{\Large Benjamin Lindner and Andr{\'e} Longtin}\\[.5em]
{\large \em  Department of Physics, University of Ottawa\\
 150 Louis Pasteur, Ottawa, KIN 6N5,  Canada}\\

\date{\today}
\end{center}
%
%
\ifcase\value{benji} \newpage \or \vskip3em \fi
\begin{abstract}
  In two recent papers, Rudolph and Destexhe (Neural Comp. {\bf 15},
  2577-2618, 2003; Neural Comp. in press, 2005) studied a leaky integrator
  model (i.e. an RC-circuit) driven by correlated (``colored'') Gaussian
  conductance noise and Gaussian current noise. In the first paper they
  derived an expression for the stationary probability density of the membrane
  voltage; in the second paper this expression was modified to cover a larger
  parameter regime. Here we show by standard analysis of solvable limit cases
  (white-noise limit of additive and multiplicative noise sources; only slow
  multiplicative noise; only additive noise) and by numerical simulations that
  their first result does not hold for the general colored-noise case and
  uncover the errors made in the derivation of a Fokker-Planck equation for
  the probability density.  Furthermore, we demonstrate analytically
  (including an exact integral expression for the time-dependent mean value of
  the voltage) and by comparison to simulation results, that the extended
  expression for the probability density works much better but still does not
  solve exactly the full colored-noise problem. We also show that at stronger
  synaptic input the stationary mean value of the linear voltage model may
  diverge and give an exact condition relating the system parameters for which
  this takes place.
\end{abstract}
\ifcase\value{benji} \newpage \or \vskip3em \fi
\section{Introduction}
The inherent randomness of neural spiking has stimulated the exploration of
stochastic neuron models for several decades \citep{Hol76,Tuc88,Tuc89}. The
subthreshold membrane voltage of cortical neurons shows strong fluctuations
{\sl in vivo} caused mainly by synaptic stimuli coming from as many as tens of
thousands of presynaptic neurons. In the theoretical literature these stimuli
have been approximated in different ways. The most biophysically realistic
description is to model an extended neuron with different sorts of synapses
distributed over the dendrite and possibly the soma, with each synapse
following its own kinetics when excited by random incoming pulses that change
the local conductance.  In a ``point-neuron'' model for the membrane potential
in the spike generating zone, this amounts to an effective conductance noise
for each sort of synapse. If the contribution of a single spike is small and
the effective input rates are high, the incoming spike trains can be well
approximated by Gaussian white noise; this is known as the diffusion
approximation of spike train input \citep[see, e.g.][]{Hol76}.  Furthermore,
these conductance fluctuations driving the membrane voltage dynamics will be
correlated in time (the noise will be ``colored'') due to the synaptic
filtering \citep{BruSer98}.  Assuming the validity of the diffusion
approximation, two further common approximations found in the theoretical
literature are (1) to replace the conductance noise by a current noise; and
(2) to neglect the correlation of the noise and to use a white noise.  To
explore the validity of these approximations has been the aim of a number of
recent theory papers \citep{RudDes03,Ric04,RicGer05,RudDes05}.\\
\citet{RudDes03} (in the following abbreviated by R\&D) recently
studied the subthreshold voltage dynamics driven by colored Gaussian
conductance and current noises, with the goal of deriving analytical
expressions for the probability density of the voltage fluctuations in the
absence of a spike-generating mechanism. Such expressions are desirable
because they permit to use experimentally measured voltage traces in vivo to
determine (or at least to obtain constraints on) synaptic parameters.  R\&D
gave a one-dimensional Fokker-Planck equation for the evolution of the
probability density of the voltage variable and solved this equation in the
stationary state. Comparing this solution to results of numerical simulations
they found a good agreement to simulations of the full model. In a recent
Note, however, they discovered a disagreement of their formula to simulations
in extreme parameter regimes.  R\&D proposed an extended expression that is
functionally equivalent to their original formula; it results from effective
correlation times that were introduced into their original formula in a
heuristic manner. According to R\&D this new expression fits simulation
results well for various parameter sets.\\
In this comment we show that both proposed formulas are not exact solutions of
the mathematical problem posed by R\&D.  We demonstrate this by the analysis
of limit cases, by means of an exact analytical result for the mean value of
the voltage as well as by numerical simulation results. The failure of the
first formula is pronounced, e.g. it fails, for instance, dramatically if the
synaptic correlation times are varied by only one order of magnitude relative
to R\&D's standard parameters. However, the extended expression, although not
an exact solution of the problem, seems to provide a reasonable approximation
for the probability density of the membrane voltage if the conductance noise
is not too strong. We also show that if the conductance noise is strong, the
model itself and not only the solutions proposed by R\&D becomes problematic:
the moments of the voltage, e.g. its stationary mean value, diverge. For the
mean value we will give an exact solution and identify by means of this
solution the parameters for which a divergence is observed.\\
This paper is organized as follows. In the next section, we introduce the
model which was studied by R\&D. In the following sections we study the limit
cases of only white noise (section \ref{sec:white_noise}), of only additive
colored noise (section \ref{sec:colored}), and of slow ("static")
multiplicative noise (section \ref{sec:static}). In section
\ref{sec:meanvalue} we derive expressions for the time-dependent and for the
stationary mean value of the voltage at arbitrary values of the correlation
times. Sec.~\ref{sec:sims} is devoted to a comparison of numerical simulations
to the various theoretical formulas. We summarize and discuss our findings in
section~\ref{sec:conclusions}. In the appendix~\ref{sec:app} we uncover the
errors in the derivation of the Fokker-Planck equation made by R\&D. We
anticipate that our results will help future investigations of the neural
colored noise problem.
%
%
\section{Basic model} 
%
%
The current balance equation for a patch of passive membrane is
\be 
\label{mem_dyn}
C_m \frac{dV(t)}{dt}=-g_L(V(t)-E_L)-\frac{1}{a}I_{syn}(t) 
\ee
where $C_m$ is the specific membrane capacity, $a$ is the membrane area, and
$g_L$ and $E_L$ the leak conductance and reversal potential, respectively.
The total synaptic current is given by
\be
I_{syn}=g_e(t)(V(t)-E_e)+g_i(t)(V(t)-E_i)-I(t) 
\ee
with $g_{e,i}$ being the noisy conductances for excitatory and inhibitory
synapses and $E_{e,i}$ the respective reversal potentials; $I(t)$ is an
additional noisy current.
With respect to the conductances, R\&D assume the diffusion approximation to
be valid. This means to approximate the superposition of incoming presynaptic
spikes at the excitatory and inhibitory synapses by Gaussian white
noise. Including a first-order linear synaptic filter, the conductances are
consequently described by Ornstein-Uhlenbeck processes (OUP); similarly, R\&D
also assume an Ornstein-Uhlenbeck process for the current $I(t)$
\ba
\label{OUP1}
\frac{dg_{e}(t)}{dt}&=&-\frac{1}{\tau_{e}}(g_{e}(t)-g_{e0})+\sqrt{\frac{2\sigma^2_{e}}{\tau_{e}}} \xi_{e}(t), \\
\label{OUP2}
\frac{dg_{i}(t)}{dt}&=&-\frac{1}{\tau_{i}}(g_{i}(t)-g_{i0})+\sqrt{\frac{2\sigma^2_{i}}{\tau_{i}}} \xi_{i}(t), \\
\label{OUP3}
\frac{d I(t)}{dt}&=&-\frac{1}{\tau_I}(I(t)-I_0)+\sqrt{\frac{2\sigma^2_I}{\tau_I}} \xi_I(t). 
\ea
Here the functions $\xi_{e,i,I}(t)$ are independent Gaussian white noise
sources with $\lr{\xi_{k}(t)\xi_{l}(t')}=\delta_{k,l}\delta(t-t')$ (
here $k,l \in \{e,i,I\}$ and the brackets $\lr{\cdots}$ stand for a
stationary ensemble average). The processes $g_e,g_i,$ and $I$ are
Gaussian distributed around the mean values $g_{e0}, g_{i0},$ and
$I_0$ with variances $\sigma_e^2,\sigma_i^2,$ and $\sigma_I^2$,
respectively
\ba
\rho_{e}(g_{e})&=&\frac{1}{\sqrt{2\pi\sigma_{e}^2}}\exp\left[-(g_{e}-g_{e0})^2/(2\sigma_{e}^2)\right], \\
\rho_{i}(g_{i})&=&\frac{1}{\sqrt{2\pi\sigma_{i}^2}}\exp\left[-(g_{i}-g_{i0})^2/(2\sigma_{i}^2)\right], \\
    \rho_{I}(I)&=&\frac{1}{\sqrt{2\pi\sigma_I^2}}\exp\left[-(I-I_0)^2/(2\sigma_I^2)\right].
  \ea 
As discussed by R\&D, these solutions permit unphysical negative conductances
which become especially important if $g_{e0}/\sigma_{e}$ and
$g_{i0}/\sigma_{i}$ are small.\\
Furthermore, the three processes are exponentially correlated with the
correlation times given by $\tau_e,\tau_i,$ and $\tau_I$, respectively
\ba
\lr{(g_{e}(t)-g_{e0})(g_{e}(t+\tau)-g_{e0})}&=&\sigma^2_{e}\exp[-|\tau|/\tau_{e}],\\
\lr{(g_{i}(t)-g_{i0})(g_{i}(t+\tau)-g_{i0})}&=&\sigma^2_{i}\exp[-|\tau|/\tau_{i}],\\
  \lr{(I(t)-I_0)(I(t+\tau)-I_0)}&=&\sigma^2_I\exp[-|\tau|/\tau_I].
\ea
Note that R\&D used another parameter to quantify the strength of the noise
processes, namely $D_{\{e,i,I\}}=2\sigma^2_{e,i,I}/\tau_{e,i,I}$.
Here we will not follow this unusual scaling\footnote{\label{fn1}In general,
two different intensity scalings for an OUP $\eta(t)$ are used in the
literature \citep[see, e.g. ][]{HanJun95}
\begin{enumerate}
\item fixing the noise intensity $Q=\int_0^\infty dT
  \lr{\eta(t)\eta(t+T)}= \sigma^2 \tau$, allowing for a proper
  white-noise limit by letting $\tau$ approach zero; with fixed noise
  intensity and $\tau\to\infty$ (static limit), the effect of the OUP
  vanishes, since the variance of the process tends to zero
\item fixing the noise variance $\sigma^2$ which leads to a finite
  effect of the noise for $\tau\to \infty$ (static limit) but makes
  the noise effect vanish as $\tau\to 0$
\end{enumerate}
R\&D use functions $\alpha_{\{e,i,I\}}(t)$, the long-time limit of which is
proportional to the noise intensity $\sigma^2_{e,i,I}\tau_{e,i,I}$.} 
but consider variations of the correlation times at either fixed variance
$\sigma^2_{e,i,I}$ of the OUPs or fixed noise intensities
$\sigma^2_{e,i,I}\tau_{e,i,I}$.\\
Eq.~(\ref{mem_dyn}) can be looked upon as a one-dimensional dynamics driven by
multiplicative and additive colored noises. Equivalently, it can be together
with \e{OUP1}, \e{OUP2}, and \e{OUP3} regarded as a four-dimensional nonlinear
dynamical system driven by only additive white noise. For such a process it is
in general quite difficult to calculate the statistics, such as the stationary
probability density $P_0(V,g_e,g_i,I)$ or the stationary marginal density for
the driven variable $\rho(V)=\int\int\int dg_e dg_i dI P_0(V,g_e,g_i,I) $
unless so-called potential conditions are met (see, e.g.  \citep{Ris84}). It
can be easily shown that the above problem does not fulfill these potential
conditions, and no solution has yet been found.\\
R\&D have proposed a solution for the stationary marginal density of the
membrane voltage $\rho(V)$ for colored noises of {\sl arbitrary correlation
times} driving their system. Their solution for the stationary probability of
the membrane voltage reads
\ba
\label{rho_rud}
\rho_{RD}(V)&=&N \exp\left[\frac{a_1}{2 b_2} \ln(b_2 V^2+b_1 V +b_0)+\right.\nz\\
&& \left. \frac{2b_2 a_0-a_1 b_1}{b_2\sqrt{4b_2 b_0-b_1^2}}
\arctan\left(\frac{2b_2 V+b_1}{\sqrt{4b_2 b_0-b_1^2}}\right)\right]
\ea
with $N$ being the normalization constant and with the abbreviations
\ba
a_0&=&\frac{1}{(C_m a)^2}(2C_m a(g_L E_La+g_{e0}E_e+g_{i0}E_i)+I_0C_ma+\sigma^2_e \tau_e E_e+\sigma^2_i \tau_i E_i),\nz\\
a_1&=&-\frac{1}{(C_m a)^2}(2C_m a(g_La+g_{e0}+g_{i0})+\sigma^2_e \tau_e+\sigma^2_i \tau_i),\nz\\
b_0&=&\frac{1}{(C_m a)^2}(\sigma^2_e \tau_e E_e^2+\sigma^2_i \tau_i E_i^2+\sigma^2_I \tau_I),\nz\\
b_1&=&-\frac{2}{(C_m a)^2}(\sigma^2_e \tau_e E_e+\sigma^2_i \tau_i E_i),\nz\\
\label{ab_rd}
b_2&=&\frac{1}{(C_m a)^2}(\sigma^2_e \tau_e+\sigma^2_i \tau_i).
\ea
In a subsequent Note on their paper \cite{RudDes05}, R\&D considered the case
of only multiplicative colored noise ($\sigma_I=0$) and showed that the
solution \e{rho_rud} does not fit numerical simulations for certain parameter
regimes. They claim that this disagreement is due to a filtering problem not
properly taken into account in their previous work. They proposed a new
solution for the case of only multiplicative noise that is functionally
equivalent to \e{rho_rud} for $\sigma_I=0$ but simply replaces correlation
times by effective correlation times
\be
\label{eff_tau}
\tau'_{e,i}=\frac{2\tau_{e,i}\tau_0}{\tau_e+\tau_0}.  
\ee 
where $\tau_0=aC_m/(ag_L+g_{e0}+g_{i0})$. Explicitely, this {\em extended
expression} is given by
\ba
\label{rho_rud2}
\rho_{RD,ext}(V)&=&N' \exp\left[ A_1
\ln\left(\frac{\sigma_e^2\tau'_e}{(C_ma)^2}
(V-E_e)^2+\frac{\sigma_i^2\tau'_i}{(C_ma)^2} (V-E_i)^2\right)+\right. \nz\\
&& \left. A_2\arctan\left(\frac{\sigma_e^2\tau'_e (V-E_e)+\sigma_i^2\tau'_i(V-E_i)}{(E_e-E_i)\sqrt{\sigma_e^2\tau'_e\sigma_i^2\tau'_i}}\right)\right]
\ea
with the abbreviations
\ba
A_1&\!\!\!=\!\!\!&-\frac{2C_ma(g_{e0}+g_{i0})+2C_ma^2 g_L+\sigma_e^2\tau'_e+\sigma_i^2\tau'_i}{2(\sigma_e^2\tau'_e+\sigma_i^2\tau'_i)}\\
A_2&\!\!\!=\!\!\!&\frac{g_La(\sigma_e^2\tau'_e (E_L-E_e)+\sigma_i^2\tau'_i(E_L-E_i))+(g_{e0}\sigma_i^2\tau'_i-g_{i0}\sigma_e^2\tau'_e
)(E_e-E_i)}{(E_e-E_i)\sqrt{\sigma_e^2\tau'_e\sigma_i^2\tau'_i}(\sigma_e^2\tau'_e+\sigma_i^2\tau'_i)/(2C_ma)}.\nz\\
\ea
The introduction of the effective correlation times was justified by
considering the effective-time constant (ETC) or Gaussian approximation from
\citet{Ric04} (see below) which reduces the system to one with additive noise.
The new formula \e{rho_rud2} fits well their simulation results for various
combinations of parameters \citep{RudDes05}.\\
In this comment, we will show that neither of these formulas yields the exact
solution of the mathematical problem. As we will show first, the original
formula fails significantly outside the limited parameter range investigated
in R\&D 2003 . Apparently, the second formula provides a good fit formula for
a number of parameter sets. It also reproduces two of the simple limit cases,
in which the first formula fails. By means of the third limit case as well as
of an exact solution for the stationary mean value (derived in
section~\ref{sec:meanvalue}, we can, however, show that the new formula is not
an exact result either.\\
To demonstrate the invalidity of the first expression in the general case, we
will show that \e{rho_rud} fails in three limits that are tractable by
standard techniques: (1) the white noise limit of all three colored noise
sources, i.e. keeping the noise intensities
$\sigma_{e,i,I}^2\tau_{e,i,I}$ fixed and letting all noise correlation
times tend to zero $\tau_{e,i}\to 0$; (2) the case of additive colored
noise only; (3) the limit of large $\tau_{e,i}$ in the case of
multiplicative colored noises with fixed variances $\sigma^2_e$ and
$\sigma^2_i$. In all cases, we also ask whether mean and variance can be
expected to be finite as it has been tacitly assumed by R\&D.\\
We will also compare both solutions proposed by R\&D as well as our own
analytical results for the limit cases to numerical simulation results.  While
the failure of the first formula \e{rho_rud} is pronounced except for a small
parameter regime, deviations of the extended expression \e{rho_rud2} are much
smaller and for six different parameter sets inspected, the new formula can be
at least regarded as a good approximation. Parameters can be found, however,
where deviations of this new formula from numerical simulations become more
serious.\\
To simplify the notation we will use the new variable $v=V-\Delta$ with
\be
\Delta=\frac{g_La E_L+g_{e0}E_e+g_{i0}E_i+I_0}{g_L a +g_{e0}+g_{i0}}.
\ee
Then the equations can be recast into 
\ba
\label{my_v}
\dot{v}&=&-\beta v-y_e(v-V_e)-y_i(v-V_i)+y_I \\
\label{my_y}
\dot{y}_{e,i,I}&=&-\frac{y_{e,i,I}}{\tau_{e,i,I}}+\sqrt{\frac{2\tilde{\sigma}^2_{e,i,I}}{\tau_{e,i,I}}} \xi_{e,i,I}(t)
\ea
with the abbreviations
\ba
\label{beta}
\beta&=&\frac{g_L a+g_{e0}+g_{i0}}{a C_m},\\
V_{e,i}&=&E_{e,i}-\Delta,\\
\label{sig_lin}
\tilde{\sigma}_{e,i,I}&=&\sigma_{e,i,I}/(aC_m). 
\ea
Once we have found an expression for the probability density of $v$,
the density for the original variable $V$ is given by the former
density taken at $v=(V-\Delta)$.\\
Finally, we briefly explain the {\em effective-time constant} (ETC) or {\em
Gaussian} approximation (cf. \citet{Ric04,RicGer05} and references therein)
which we will refer to later on. Assuming weak noise sources, the voltage will
fluctuate around the deterministic equilibrium value $v=0$ with an amplitude
proportional to the square root of the sum of the noise variances, e.g. for
only excitatory conductance fluctuations we would have a proportionality to
the standard deviation of $y_e$, i.e.  $\lr{|v|}\propto\sqrt{\lr{y_e^2}}$.  From
this we can see that the multiplicative terms $y_e V$ and $y_I V$ make a
contribution proportional to the squares of the standard deviations and can
therefore be neglected for weak noise. The resulting dynamics contains only
additive noise sources
\be
\label{eff_time_constant}
\dot{v}=-\beta v+y_e V_e+y_i V_i+y_I.  
\ee
The stationary probability density is a Gaussian
\be
\label{rho_etc}	
\rho_{ETC}(v)=\frac{\exp[-v^2/(2\lr{v^2}_{ETC})]}{\sqrt{2\pi \lr{v^2}_{ETC}}}
\ee
 with zero mean and a variance given by \citep{Ric04}
\be
\label{var_etc} \lr{v^2}_{ETC}=V_e^2 \frac{\sigma_e^2\tau_e/\beta}{1+\beta\tau_e}+V_i^2
\frac{\sigma_i^2\tau_i/\beta}{1+\beta\tau_i}
+\frac{\sigma_I^2\tau_I/\beta}{1+\beta\tau_I} \ee
The solution takes into account the effect of the
mean conductances on the effective membrane time constant $1/\beta$
through \e{beta}.
\section{The white-noise limit}
\label{sec:white_noise}
If we fix the noise intensities
\be
\label{noise_int}
Q_{e,i,I}=\tilde{\sigma}_{e,i,I}^2\tau_{e,i,I},
\ee
we may consider the limit of white noise by letting
$\tau_{e,i,I}\to 0$. A special case of this has been recently
considered by \citet{Ric04} with $\sigma_I=0$ (only multiplicative
noise is present).\\
In the white-noise limit, the three OUPs approach mutually independent
white-noise sources
\be
y_e\to \sqrt{2Q_e} \xi_e(t),\;\;y_i\to \sqrt{2Q_i} \xi_i(t),\;\;y_I\to \sqrt{2Q_I} \xi_I(t) 
\ee
and, thus the current balance equation \e{my_v} becomes
\be
\label{v_wn0}
\dot{v}=-\beta v-\sqrt{2Q_e}(v-V_e) \xi_e(t)-\sqrt{2Q_i}(v-V_i)\xi_i(t)+ \sqrt{2Q_I} \xi_I(t) 
\ee
which is equivalent\footnote{The sum of three independent Gaussian noise sources
gives one Gaussian noise the variance of which equals the sum of the variances 
of the single noise sources.} to a driving by a single Gaussian noise $\xi(t)$
\be
\label{v_wn}
\dot{v}=-\beta v+\sqrt{2Q_e (v-V_e)^2 +2Q_i(v-V_i)^2+2Q_I} \xi(t) 
\ee
with $\lr{\xi(t)\xi(t')}=\delta(t-t')$. Since we approach the white
noise limit with having in mind colored noises with negligible
correlation times, \e{v_wn} has to be interpreted in the sense of
Stratonovich \citep[see sec. 6.5]{Ris84,Gar85}. The drift and diffusion coefficients then
read \citep[cf. (3.95) therein]{Ris84}
\ba D^{(1)}&=&-\beta v +Q_e (v-V_e)+Q_i (v-V_i)=-\beta
v+\frac{1}{2}\frac{d
  D^{(2)}}{dv} \\
D^{(2)}&=&Q_I+Q_e (v-V_e)^2+Q_i (v-V_i)^2 
\ea
and the stationary solution of the probability density is given by
\citep[cf. eq.~(5.13) therein]{Ris84} 
\ba \rho_{wn}(v)&=& N \exp\left[-\ln(D^{(2)})+\intl^v dx
  \frac{D^{(1)(x)}}{D^{(2)(x)}}\right] 
\ea
where the subscript "wn" refers to white noise.\\
After carrying out the integral, the solution can be written as
follows 
\ba
\label{rho_wn}
\rho_{wn}(v)&=& N
\exp\left[-\frac{\beta+\tilde{b}_2}{2\tilde{b}_2}\ln(\tilde{b}_2v^2+\tilde{b}_1v+\tilde{b}_0)+\right.\nz\\
&& \left.\frac{\beta \tilde{b}_1}{\tilde{b}_2\sqrt{4\tilde{b}_0\tilde{b}_2-\tilde{b}_1^2}} \arctan\left(\frac{2\tilde{b}_2v+\tilde{b}_1}{\sqrt{4\tilde{b}_0\tilde{b}_2-\tilde{b}_1^2}}\right)\right]
\ea
with the abbreviations 
\ba
\label{lins_bee0}
\tilde{b}_0&=&Q_I+Q_e V_e^2+Q_i V_i^2\\
\label{lins_bee1}
\tilde{b}_1&=&-2(Q_e V_e+Q_i V_i)\\
\label{lins_bee2}
\tilde{b}_2&=&Q_e+Q_i 
\ea
Different versions of the white-noise case have been discussed and also
analytically studied in the literature \citep[see,
e.g.,][]{HanTuc83,LanLan87,LanLan94,Ric04}. In particular, eq.~(\ref{rho_wn})
is consistent with the expression for the voltage density in a leaky
integrate-and-fire neuron driven by white noise\footnote{The density \e{rho_wn} results from
eq.~(18) in \citet{Ric04} when firing and reset in the integrate-and-fire
neuron become negligible. This can be formally achieved by letting threshold
and reset voltage go to positive infinity.} given by
\citet{Ric04}.\\
Since \e{rho_rud} was proposed by R\&D as the solution for the probability
density at arbitrary correlation times of the colored-noise sources, it should
be also valid in the white-noise limit and agree with \e{rho_wn}. On closer
inspection it becomes apparent that both \e{rho_rud} and \e{rho_rud2} have the
structure of the white-noise solution \e{rho_wn}. Comparing the factors of the
terms in the exponential, we find that the first solution (in terms of the
shifted voltage variable and using the noise intensities
\e{noise_int}) can be written as follows
\be
\label{rel_rho_rud_wn}
\rho_{RD}(v,Q_e,Q_i,Q_I)=\rho_{wn}(v,Q_e/2,Q_i/2,Q_I/2),
\ee
where the additional arguments of the functions indicate the parametric
dependence of the densities on the noise intensities. According to
\e{rel_rho_rud_wn}, if formulated in terms of the noise intensities (and not
the noise variances) the first formula proposed by \rd does not depend on the
correlation times $\tau_{e,i,I}$ at all. {\em Furthermore, it is evident from 
\e{rel_rho_rud_wn} that the expression is incorrect in the white noise limit.}
If all correlation times $\tau_{e,i,I}$ simultaneously go to zero, the
density approaches the white-noise solution with only half of the true values
of the noise intensities. The density will certainly depend on the noise
intensities and will change if one uses only half of their values.\\
We may also rewrite R\&D's extended expression \e{rho_rud2} in terms of the
white-noise density
\be
\label{rel_rho_rud_wn2}
\rho_{RD,ext}(v,Q_e,Q_i)=\rho_{wn}(v,Q_e/(1+\beta\tau_e),Q_i/(1+\beta\tau_i),Q_I=0)
\ee
This solution agrees with the original solution by R\&D only for the specific
parameter set
\be
\tau_e=\tau_i=1/\beta.
\ee
We note that since the extended solution can be expressed by means of the
white-noise density it makes sense to describe the extended expression by
means of effective noise intensities
\be
\label{eff_noise}
Q'_{e,i}=\frac{Q_{e,i}}{1+\beta \tau_{e,i}} 
\ee
rather than in terms of the effective correlation times $\tau'_{e,i}$
(cf. \e{eff_tau}) used by R\&D. The assertion behind \e{rel_rho_rud_wn2} is
the following: the probability density of the membrane voltage is always
equivalent to the white-noise density; correlations in the synaptic input
(i.e. finite values of $\tau_{e,i,I}$) just lead to rescaled (smaller)
noise intensities $Q'_{e,i}$ given in \e{eff_noise}.\\
If we consider the white-noise limit of the r.h.s. of
\e{rel_rho_rud_wn2}, we find that the extended expression \e{rho_rud2}  reproduces this
limit, i.e.
\be 
\liml_{\tau_e,\tau_i\to 0}
\rho_{RD,ext}(V,Q_e,Q_i)=\rho_{wn}(V,,Q_e,Q_i,Q_I=0). 
\ee
So there is no problem with the extended expression in the white-noise
limit.\\
\subsection{Divergence of moments in the white-noise limit and in R\&D's
expressions for the probability density}
We consider the density \e{rho_wn} in the limits $v\to\pm\infty$ and conclude
whether the moments and, in particular, the mean value of the white-noise
density are finite; similar arguments will be applied to the solutions
proposed by R\&D.\\
At large $v$ and to leading order in $1/v$ we obtain
\be
\label{rho_wn_asymp}
\rho_{wn}(v)\sim |v|^{\te-\frac{\beta+\tilde{b}_2}{\tilde{b}_2}} N
\tilde{b}_2^{\te-\frac{\beta+\tilde{b}_2}{2\tilde{b}_2}}
\exp\left[\pm\frac{\beta
    \tilde{b}_1}{\tilde{b}_2\sqrt{4\tilde{b}_0\tilde{b}_2-\tilde{b}_1^2}}
  \frac{\pi}{2}\right]\;\;\; \mbox{as}\;\;\; v\to\pm\infty. 
\ee
When calculating the $n$th moment, we have to multiply with $v^n$ and obtain a
non-diverging integral only if $v^n \rho_{wn}(v)$ decays faster than
$v^{-1}$. This is the case only if $n-(\beta+\tilde{b}_2)/\tilde{b}_2<-1$ or
using \e{lins_bee2}
\be
\label{condi_n_wn}
|\lr{v^n}_{wn}|<\infty \;\; \mbox{iff}\;\; \beta>n(Q_e+Q_i)
\ee 
where ``iff'' stands for ``if and only if'' and the index ``wn''
indicates that we consider the white-noise case.  Note that no
symmetry argument applies for odd $n$ since the asymptotic limits
differ for $\infty$ and $-\infty$ according to \e{rho_wn_asymp}. For
the mean, this implies that
\be
\label{condi_wn}
|\lr{v}_{wn}|<\infty \;\; \mbox{iff}\;\; \beta>Q_e+Q_i
\ee
otherwise the integral diverges.\\
In general, the power law tail in the density is a hint that (for
white noise at least) we face the problem of rare strong deviations in
the voltage, that are due to the specific properties of the model
(multiplicative Gaussian noise).\\
Because of \e{rel_rho_rud_wn} similar conditions (differing by a
prefactor of 1/2 on the respective right hand sides) also apply for
the finiteness of the mean and variance of the original solution
\e{rho_rud} proposed by R\&D. For the mean value of this solution one obtains
the condition
\be
\label{condi_rud}
|\lr{v}_{RD}|<\infty \;\; \mbox{iff}\;\;\beta>\frac{Q_e+Q_i}{2},
\ee
which should hold true in the general colored noise case but does not agree 
with the condition in \e{condi_wn} even in the white-noise case.\\
From the extended expression we obtain
\be
\label{condi_rud2}
|\lr{v}_{RD,ext}|<\infty \;\; \mbox{iff}\;\;\beta>\frac{Q_e}{1+\beta\tau_e}  +\frac{Q_i}{1+\beta\tau_i}.
\ee
Note that \e{condi_rud2} agrees with \e{condi_wn} only in the white-noise case
(i.e. for $\tau_e,\tau_i\to 0$). Below we will show that \e{condi_wn} gives
the correct condition for a finite mean value in the general case of arbitrary
correlation times, too. Since for finite $\tau_e,\tau_i$, the two conditions
\e{condi_wn} and \e{condi_rud2} differ, we can already conclude that the \e{rho_rud2}
that led to condition \e{condi_rud2} cannot be the exact solution of the
original problem.
\section{Additive colored noise}
\label{sec:colored}
Setting the multiplicative colored noise sources to zero, R\&D obtain an
expression for the marginal density in case of additive colored noise only
(cf. eq.~(3.7-3.9) in R\&D)
\be
\label{rho_rud_add1}
\rho_{add,RD}(V)=N \exp\left[-\frac{a^2 g_L C_m(V-E_L-I_0/(g_L a))^2}{\sigma^2_I \tau_I}\right] 
\ee
which corresponds in our notation and in terms of the shifted variable 
$v$ to 
\be
\label{rho_rud_add2}
\tilde{\rho}_{add,RD}(v)=N \exp\left[-\frac{\beta v^2}{Q_I}\right] \ee 
Evidently, once more a factor 2 is missing already in the white-noise
case (where the process $v(t)$ itself becomes an Ornstein-Uhlenbeck
process), since for an OUP we should have $\rho\sim\exp[-\beta
v^2/(2Q_I)]$. However, there is also a missing additional dependence
on the correlation time.\\
For additive noise only, the original problem given in \e{mem_dyn}
reduces to
\ba
\label{add_v}
\dot{v}&=&-\beta v+y_I, \\
\label{add_y}
\dot{y_I}&=&-\frac{1}{\tau_I} y_I+\frac{\sqrt{2Q_I}}{\tau_I} \xi_I(t).
\ea
This system is mathematically similar to the Gaussian approximation or
effective-time constant approximation \e{rho_etc} in which also no multiplicative
noise is present. The density function for the voltage is well known;
for the sake of clarity we show here how to calculate it.\\
The system eqs.~(\ref{add_v},\ref{add_y}) obeys the two-dimensional
Fokker-Planck equation
\be
\label{P_vy}
\partial_t P(v,y_I,t)=\left[\partial_v(\beta v-y_I)+\partial_{y_I}\left(\frac{y_I}{\tau_I}+\frac{Q_I}{\tau^2_I}\partial_{y_I}\right)\right]P(v,y_I,t)
\ee
The stationary problem ($\partial_t P_0(v,y_I)=0$) is solved by an
ansatz $P_0(v,y)\sim \exp[Av^2+Bvy+C y^2]$ yielding the solution for
the full probability density
\be
\label{rho_vy}
 P_0(v,y_I)=N
\exp\left[\frac{c}{2}\left(y_I^2-2\beta
    vy_I-\frac{Q_I\beta}{\tau^2_I}c
    v^2\right)\right],\;\;c=-\frac{\tau_I(1+\beta\tau_I)}{Q_I} 
\ee
Integrating over $y_I$, yields the correct marginal density 
\be
\label{rho_lin_add}
\rho_{add}(v)=\sqrt{\frac{\beta(1+\beta\tau_I)}{2\pi Q_I}}
\exp\left[-\frac{\beta v^2}{2Q_I}(1+\beta\tau_I)\right] 
\ee 
which is in disagreement with \e{rho_rud_add2} and hence also with
\e{rho_rud_add1}. From the correct solution given in \e{rho_lin_add},
we also see what happens in the limit of infinite $\tau$ for fixed
noise intensity $Q_I$: the exponent tends to minus infinity except at
$v=0$ or, put differently, the variance of the distribution tends to
zero and we end up with a $\delta$ function at $v=0$. This limit makes
sense (cf. footnote \ref{fn1}) but is not reflected at all in the
original solution \e{rho_rud2} given by R\&D.\\
We can also rewrite the solution in terms of the white-noise solution
in the case of vanishing multiplicative noise 
\be
\label{rho_lin_add_rel}
\rho_{add}(v)=\rho_{wn}(v,Q_e=0,Q_i=0,Q_I/[1+\beta\tau_I]).
\ee 
Thus, for the additive noise is true, what has been assumed by R\&D in the
case of multiplicative noise: the density in the general colored-noise case is
given by the white-noise density with a rescaled noise intensity
$Q'_I=Q_I/[1+\beta\tau_I]$ (or equivalently, rescaled correlation time
$\tau'_I=2\tau_I/[1+\beta\tau_I]$ in \e{rho_rud_add2} with
$Q_I=\sigma^2\tau_I'$).\\
We cannot perform the limit of only additive noise in the extended expression
\e{rho_rud2} proposed by R\&D because this solution was meant for the case of only
multiplicative noise. If, however, we generalize the solution \e{rho_rud2} to
the case of additive and multiplicative colored noises, we can consider the
limit of only additive noise in this solution. This is done by taking the
original solution by R\&D \e{rho_rud} and replacing not only the correlation
times of the multiplicative noises $\tau_{e,i}$ by the effective ones
$\tau'_{e,i}$ but also that of the additive noise $\tau_I$ by an effective
correlation time
\be
\tau'_I=\frac{2\tau_I}{1+\tau_I \beta}. 
\ee
If we now take the limit $Q_e=Q_i=0$, we obtain the correct density
\be
\label{rho_lin_add_ext}
\rho_{rud,ext,add}(v)=\rho_{wn}(v,Q_e=0,Q_i=0,Q_I/[1+\beta\tau_I])
\ee
as becomes evident on comparing the r.h. sides of \e{rho_lin_add_ext} and
\e{rho_lin_add_rel}.\\
Finally, we note that the case of additive noise is the only limit
that does not pose any condition on the finiteness of the moments.
\section{Static multiplicative noises only (limit of large $\tau_{e,i}$)}
\label{sec:static}
Here we assume for simplicity $\tilde{\sigma}_I=0$ and consider multiplicative
noise with fixed variances $\tilde{\sigma}^2_{e,i}$ only. If the noise
sources are much slower than the internal time scale of the system, i.e. if
$1/(\beta\tau_e)$ and $1/(\beta\tau_i)$ are practically zero, we can neglect
the time derivative in \e{my_v}. This means that the voltage adapts
instantaneously to the multiplicative ("static") noise sources which is
strictly justified only for $\beta\tau_e,\beta \tau_i\to\infty$. If
$\tau_e,\tau_i$ attain large but finite values ($\beta\tau_i,\beta\tau_i\gg
1$), the formula derived below will be an approximation that works the better
the larger these values are. Because of the slowness of the noise sources
compared to the internal time scale, we call the resulting expression the
"static-noise" theory for simplicity. This {\em does not} imply that the total
system (membrane voltage plus noise sources) is not in the stationary state:
we assume that any initial condition of the variables has decayed on a time
scale $t$ much larger\footnote{In the strict limit of $\beta\tau_e,\beta
\tau_i\to\infty$ this would imply that $t$ goes stronger to infinity than the
correlation times $\tau_{e,i}$ do.} than $\tau_{e,i}$.  For a simulation of
the density this has the practical implication that we should choose a
simulation time much larger than any of the involved correlation times.\\
Setting the time derivative in \e{my_v} to zero, we can determine at which
position the voltage variable will be for a given quasi-static pair of
$(y_e,y_i)$ values, yielding
\be
v=\frac{y_e V_E +y_i V_i}{\beta+y_e+y_i}
\ee
This sharp position will correspond to a $\delta$ peak of the
probability density
\be
\delta\left(v-\frac{y_eV_E +y_i V_i}{\beta+y_e+y_i}\right)=\frac{|y_i(V_i-V_e)-\beta V_e|}{(v-V_e)^2}\delta\left(y_e+\frac{\beta v+y_i(v-V_i)}{(v-V_e)}\right)
\ee
(here we have used $\delta(ax)=\delta(x)/|a|$).  This peak has to be
averaged over all possible values of the noise, i.e. integrated over
the two Gaussian distributions in order to obtain the marginal density
\ba
\rho_{static}(v)&=&\lr{\delta(v-v(t))}\nz\\
&=&\intl_{-\infty}^\infty\intl_{-\infty}^\infty \frac{dy_e dy_i}{2\pi \tilde{\sigma}_i \tilde{\sigma}_e}\frac{|y_i(V_i-V_e)-\beta V_e|}{(v-V_e)^2}
\delta\left(y_e+\frac{\beta v+y_i(v-V_i)}{(v-V_e)}\right)\times\nz\\
&&\hspace{1cm} \exp\left[-\frac{y_e^2}{2\tilde{\sigma}_e^2} -\frac{y_i^2}{2\tilde{\sigma}_i^2}\right] 
\ea
Carrying out these integrals yields
\be
\label{rho_static}
\rho_{static}(v)=\frac{\tilde{\sigma}_e\tilde{\sigma}_i |V_e-V_i|}{\pi\beta^2\mu(v)}e^{-\te\frac{v^2}{2\mu(v)}}\left[e^{-\te\frac{\nu(v)}{\mu(v)}}+\sqrt{\frac{\pi \nu(v)}{\mu(v)}} \mbox{erf}\left(\sqrt{\frac{ \nu(v)}{\mu(v)}}\right)\right]
\ee
where erf$(z)$ is the error function \citep{AbrSte70} and the functions
$\mu(v)$ and $\nu(v)$ are given by
\ba
\mu(v)&=&\frac{\tilde{\sigma}_e^2(v-V_e)^2+\tilde{\sigma}_i^2(v-V_i)^2}{\beta^2}\\
\nu(v)&=&\frac{[\tilde{\sigma}_e^2V_e(v-V_e)+\tilde{\sigma}_i^2V_i(v-V_i)]^2}{2\tilde{\sigma}_e^2\tilde{\sigma}_i^2(V_e-V_i)^2}
\ea
If one of the expressions by R\&D \e{rho_rud} or \e{rho_rud2} would be the
correct solution, it should converge for $\sigma_I=0$ and
$\tau_{e,i}\to\infty$ to the formula for the static case \e{rho_static}.
In general, this is not the case since the functional structure of the
white-noise solution and that of the static-noise solution are quite
different. There is, however, one limit case in which the extended expression
yields the same (although trivial) function. If we fix the noise intensities
$Q_{e,i}$ and let the correlation times go to infinity, the variances will go
to zero and the static-noise density \e{rho_static} approaches a $\delta$ peak
at $v=0$.  Although the extended expression
\e{rho_rud2} has a different functional dependence on system parameters and
voltage, the same thing happens in the extended formula for $\tau_{e,i}\to
\infty$ because the effective noise intensities
$Q'_{e,i}=Q_{e,i}/(1+\beta\tau_{e,i})$ approach zero in this limit. The
white-noise solution at vanishing noise intensities is, however, also a
$\delta$ peak at $v=0$. {\em Hence, in the limit of large correlation time at
fixed noise intensities, both the static-noise theory \e{rho_static} and the
extended expression yield both the probability density of a noise-free system
and therefore agree.} For fixed variance where a non-trivial large-$\tau$
limit of the probability density exists, the static-noise theory and the
extended expression by R\&D differ as we will also numerically verify later
on.\\
A final remark concerns the asymptotic behavior of the static-noise solution
\e{rho_static}.  The asymptotic expansions for $v\to\pm\infty$ show that the
density goes like $|v|^{-2}$ in either limits. Hence, in this case we cannot
obtain a finite variance of the membrane voltage at all (the integral $\int
dv\; v^2 \rho_{static}(v)$ will diverge). The mean may be finite since the
coefficients of the $v^{-2}$ term are symmetric in $v$. The estimation in the
following section, however, will demonstrate that this is valid only strictly
in the limit $\tau_{e,i}\to\infty$ but not at any large but finite value of
$\tau_{e,i}$. So the mean may diverge for large but finite $\tau_{e,i}$.
\section{Mean value of the voltage for arbitrary values of the correlation times}
\label{sec:meanvalue}
By inspection of the limit cases we have already seen that the moments
do not have to be finite for an apparently sensible choice of
parameters.  For the white-noise case it was shown that the mean of
the voltage is finite only if $\beta>Q_e+Q_i$.\\
Next, we show by direct analytical solution of the stochastic
differential equation \e{my_v} involving the colored noise sources
\e{my_y} that this condition (i.e. \e{condi_wn}) holds in general and
thus a divergence of the mean is obtained for $\beta<Q_e+Q_i$.\\
For only one realization of the process \e{my_v}, the driving
functions $y_e(t)$, $y_i(t)$, and $y_I(t)$ can be regarded as just
time-dependent parameters in a linear differential equation. The
solution is then straightforward (see also \citet{Ric04} for the
special case of only multiplicative noise)
\ba
\label{sol_direct}
v(t)&=&v_0 \exp\left[-\beta t-\intl_0^t du(y_e(u)+y_i(u))\right]+\nz\\
&&\hspace{-3.5em} \intl_0^t ds (V_e y_e(s)+V_i y_i(s)+y_I(s)) e^{-\beta(t-s)}\exp\left[-\intl_s^t du(y_e(u)+y_i(u))\right]
\ea
The integrated noise processes $w_{e,i}(s,t)=\intl_s^t du
y_{e,i}(u)$ in the exponents are independent Gaussian processes
with variance
\be
\label{var_int}
\lr{w^2_{e,i}(s,t)}=2Q_{e,i}(t-s-\tau_{e,i}+\tau_{e,i}e^{-(t-s)/\tau_{e,i}}).
\ee
For a Gaussian variable we know that $\lr{e^{w}}=e^{\lr{w^2}/2}$
\citep{Gar85}. Using this relation for the integrated noise processes
together with \e{var_int} and expressing the average
$\lr{y_{e,i}(s)\exp[-\intl_s^t du y_{e,i}(u)]}$ by a
derivative of the exponential with respect to $s$, we find an integral
expression for the mean value
\ba
\label{mean_direct}
\lr{v(t)}&=&v_0 e^{(Q_e+Q_i-\beta)t}\exp\left[-\tau_e f_e(t) -\tau_i f_i(t)\right]\nz\\
 && -\intl_0^t ds \{V_e f_e(s) +V_i f_i(s)\}e^{(Q_e+Q_i-\beta)s-\tau_e f_e(s)-\tau_if_i(s)}
\ea
where $f_{e,i}(s)=Q_{e,i}(1-\exp[-s/\tau_{e,i}])$.  The stationary mean value
corresponding to the stationary density is obtained from this expression in
the asymptotic limit $t\to\infty$.  We want to draw the reader's attention to
the fact that this mean value is finite exactly for the same condition as for
the white noise case, i.e. for
\be
\label{condi_colored}
|\lr{v}|<\infty \;\; \mbox{iff}\;\; \beta>Q_e+Q_i
\ee
Firstly, this is so because otherwise the exponent $(Q_e+Q_i-\beta)t$
in the first line is positive and the exponential diverges for
$t\to\infty$. Furthermore, if $\beta<Q_e+Q_i$ the exponential in the
integrand diverges at large $s$.\\
In terms of the original parameters of R\&D the condition for a finite
stationary mean value of the voltage reads
\be
\label{condi_rud_param}
|\lr{v}|<\infty \;\; \mbox{iff}\;\;
g_La+g_{e0}+g_{i0}>\frac{\sigma_e^2\tau_e+\sigma_i^2 \tau_i}{aC_m} 
\ee
Note that this depends also on $a$ and $C_m$, and not only on the synaptic
parameters. R\&D use as standard parameter values \citep[p.  2589]{RudDes03}
$g_{e0}=0.0121$ $\mu$S, $g_{i0}=0.0573$ $\mu$S, $\sigma_{e}=0.012$ $\mu$S,
$\sigma_{i}=0.0264$ $\mu$S, $\tau_{e}=2.728$ ms, $\tau_{i}=10.49$ ms,
$a=34636\mu \mbox{m}^2$, and $C_m=1 \mu$F/cm$^2$. They state that the
parameters have been varied in numerical simulations from 0\% to 260\%
relative to these standard values covering more than "the physiological range
observed in vivo" \citep{RudDes03}. Inserting the standard values into the
relation \e{condi_rud_param} yields 
\be
 0.0851\mu\mbox{S}>0.0221\mu\mbox{S}.
\ee 
So in this case the mean will be finite. However, using twice of the standard
value for the inhibitory noise standard deviation, i.e.  $\sigma_i=0.0528$
$\mu$S (corresponding to 200\% of the standard value) and all other parameters
as before, leads already to a diverging mean because we obtain
$0.0852\mu\mbox{S}$ on the right hand side of \e{condi_rud_param} while the
left hand side is unchanged.  This means, even in the parameter regime that
R\&D studied, the model predicts an infinite mean value of the voltage. A
stronger violation of \e{condi_rud_param} will be observed by either
increasing the standard deviations $\sigma_{e,i}$ and/or correlation times
$\tau_{e,i}$ or by decreasing the mean conductances $g_{e,i}$. We also note
that for higher moments, and especially for the variance, the condition for
finiteness will be even more restrictive as can be concluded from the limit
cases investigated before.\\
The stationary mean value at arbitrary correlation times can be inferred from
\e{mean_direct} by taking the limit $t\to\infty$.  Assuming the relation
\e{condi_colored} holds true, we can neglect the first term involving the initial
condition $v_0$ and obtain
\be
\label{mean_stat_exact}
\lr{v}=-\intl_0^\infty ds \{V_ef_e(s)+V_if_i(s)\}
\exp[(Q_e+Q_i-\beta)s-\tau_e f_e(s)-\tau_i f_i(s)]
\ee
We can also use \e{mean_stat_exact} to recover the white-noise result
for the mean as for instance found in \cite{Ric04} by taking
$\tau_{e,i}\to 0$. In this case we can integrate
\e{mean_stat_exact} and obtain 
\ba
\label{mean_stat_wn}
\lr{v}_{wn}&=&-\{V_eQ_e+V_iQ_i \} \intl_0^\infty ds 
\exp[(Q_e+Q_i-\beta)s] \nz\\
&=&-\frac{V_eQ_e+V_iQ_i }{\beta -Q_e-Q_i}
\ea
Because of the similarity of the R\&D solution to the white-noise
solution (cf. \e{rel_rho_rud_wn}), we can also infer that the mean
value of the former density is \be
\label{mean_stat_rud}
\lr{v}_{RD}=-\frac{V_eQ_e+V_iQ_i }{2\beta -Q_e-Q_i}.
\ee
Note the different prefactor of $\beta$ in the denominator that is due to the
factor $1/2$ in noise intensities of the solution \e{rho_rud} by R\&D.\\
Finally, we can also determine easily the mean value for the extended solution
by R\&D \citep{RudDes05} since also this solution is equivalent to the
white-noise solution with rescaled noise intensities.  Using the noise
intensities $Q'_{e,i}$ from
\e{eff_noise} we obtain 
\ba
\label{mean_stat_rud_ext}
\lr{v}_{RD,ext}&=&-\frac{V_e Q'_e(\tau_e)+V_iQ'_i(\tau_i) }{\beta
-Q'_e(\tau_e)-Q'_i(\tau_i)}\nz\\
&=&-\frac{V_e Q_e(1+\beta\tau_i)+V_iQ_i(1+\beta\tau_e) }{\beta(1+\beta\tau_i)(1+\beta\tau_e)
-Q_e(1+\beta\tau_i)-Q_i(1+\beta\tau_e)}
\ea
We will verify numerically that this expression is not equal to the
exact solution \e{mean_stat_exact}. One can, however, show that for
small to medium values of the correlation times $\tau_{e,i}$ and weak
noise intensities these differences are not drastic. If we expand both
\e{mean_direct} and \e{mean_stat_rud_ext} for small noise intensities
$Q_e,Q_i$ (assuming for the former that the products $Q_e\tau_e ,Q_i\tau_i$
are small, too), the resulting expressions agree to first order and also agree
with a recently derived weak noise result for filtered Poissonian shot noise
given by \citet[cf. eq.(D.3)]{RicGer05}
\ba 
\lr{v}_{RD,ext}&\approx & \lr{v}\approx -\frac{V_e
  Q_e(1+\beta\tau_i)+V_iQ_i(1+\beta\tau_e)
}{\beta(1+\beta\tau_i)(1+\beta\tau_e)}+{\cal O}(Q_e^2,Q_i^2) 
\ea
The higher order terms differ and that is why a discrepancy between
both expressions can be seen at non-weak noise.\\
The results for the mean value achieved in this section are useful in two
respects.  Firstly, we can check whether trajectories indeed diverge for
parameters where the relation \e{condi_colored} is violated. Secondly, the
exact solution for the stationary mean value and the simple expressions
resulting for the different solutions proposed by R\&D can be compared in
order to reveal their range of validity. This is done in the next section.
\section{Comparison to simulations}
\label{sec:sims}
Here we compare the different formulas for the probability density of the
membrane voltage and its mean value to numerical simulations for different
values of the correlation times, restricting ourselves to the case of
multiplicative noise only. We will first discuss the original expression
\e{rho_rud} proposed by R\&D and the analytical solutions for the limit cases
of white and static multiplicative noise (\e{rho_wn} and \e{rho_static},
respectively); later we examine the validity of the new extended
expression. Finally, we will also check the stationary and time-dependent mean
value of the membrane voltage and discuss how well these simple statistical
characteristics are reproduced by the different theories including our exact
result \e{mean_direct}.\\
To check the validity of the different expressions we will use first a
dimensionless parameter set where $\beta=1$ but also the original parameter
set used by R\&D (2003). In both cases we consider variations of the
correlation times between three orders of magnitude (i.e.  standard values are
varied between 10\% and 1000\%). Note that the latter choice goes beyond the
range originally considered by R\&D (2003) where parameter variations were
limited to the range 0\%-260\%.  

\subsection{Probability density of the membrane voltage - original expression
by R\&D}
In a first set of simulations we ignore the physical dimensions of all the
parameters and pick rather arbitrary but simple values ($\beta=1, Q_i=0.75,
Q_e=0.075$).  Keeping the ratio of the correlation times ($\tau_I=5\tau_e$)
and the values of the noise intensities $Q_e, Q_i$ fixed, we vary the
correlation times. In \bi{simu_theo} simulation results are shown for
$\tau_e=10^{-2},10^{-1},1,$ and $10$. We recall that with a fixed noise
intensity according to the result by R\&D given in \e{rho_rud}, the
probability should not depend on $\tau_e$ at all.\\
\begin{figure}[h!]
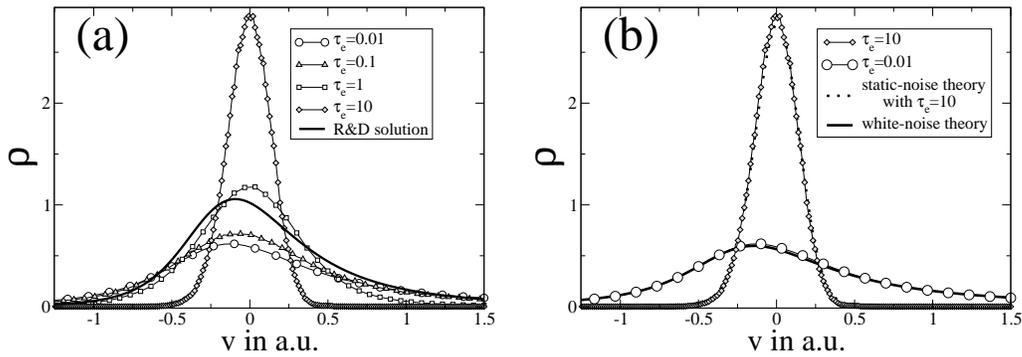

\centerline{\parbox{7cm}{\epsfig{file=fig1a.eps,width=6.5cm,angle=0}}\parbox{7.cm}{\epsfig{file=fig1b.eps,width=6.5cm,angle=0}}}
\caption{\small 
 Theoretical expressions for the probability density of the shifted
 membrane voltage $v=V-\Delta$ compared to results of numerical
 simulations in case of pure multiplicative noises.  Panel a: The
 density according to \e{rho_rud} (theory by R\&D) is compared to
 simulations at different correlation times ($\tau_e$ as indicated,
 $\tau_i=5\tau_e$). Since the noise intensities are fixed ($Q_e=0.075,
 Q_i=0.75$), the simulated densities at different correlation times
 should fall all onto the solid line according to \e{rho_rud} which is
  not the case. Panel b: The simulations at small
 ($\tau_e=0.01$) and large ($\tau_e=10$) correlation times are
 compared to our expressions found in the limit case of white and
 static noise, i.e. \e{rho_wn} and \e{rho_static}, respectively.  Note
 that, in the constant-intensity scaling, \e{rho_static} depends
 implicitly on $\tau_{e,i}$ since the variances change as
 $\sigma_{e,i}=Q_{e,i}/\tau_{e,i}$.  In both panels
 $\beta=1$ and $Q_I=0$.  For the simulations here and in the following
 figures, we followed a single realization $v(t)$ using a simple Euler
 procedure. We used a time step of $\Delta t=0.001$ and a simulation
 time of $10^{8}$ time steps. The probability density at a certain
 voltage is then proportional to the time spent by the realization in
 a small region around this voltage. Decreasing $\Delta t$ or
 increasing the simulation time did not change the above results.
\label{fig:simu_theo}}
\end{figure}

\noindent
It is obvious, however, in \bi{simu_theo}a that the simulation data depend
strongly on the correlation times in contrast to what is predicted by
\e{rho_rud}. The difference between the original theory by R\&D and the
simulations is smallest for an intermediate correlation time ($\tau_e=1$). In
contrast to the general discrepancy between simulations and \e{rho_rud}, the
white-noise formula \e{rho_wn} and the formula from the static-noise theory
(cf. solid and dotted lines in \bi{simu_theo}b) agree well with the
simulations at $\tau_e=0.01$ (circles) and $\tau_e=10$ (diamonds),
respectively.  The small differences between simulations and theory decrease
as we go to smaller or larger correlation times, respectively, as expected.\\
R\&D also present results of numerical simulations \citep{RudDes03}.  These
simulations seem to agree fairly well with their formula. In order to give a
flavor of the reliability of these data we have repeated the simulations for
one parameter set in
\citep[Fig.2b]{RudDes03}. These data are shown in \bi{simu_theo2}b and
compared to the original solution \e{rho_rud} by R\&D.\\
\begin{figure}[h!]
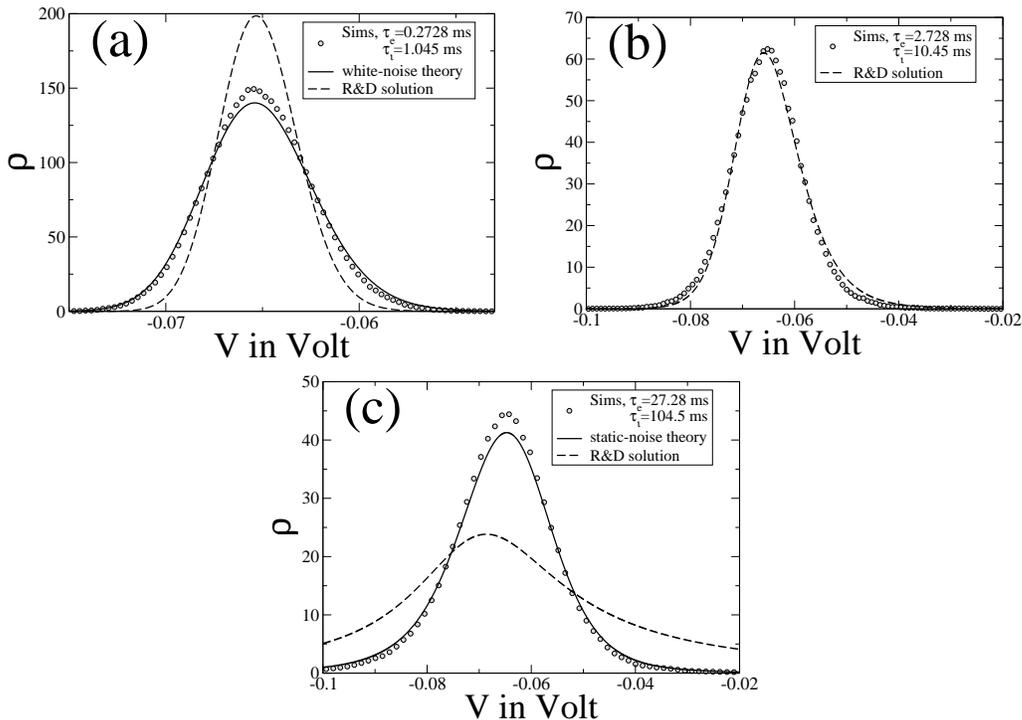

\centerline{\parbox{7.cm}{\epsfig{file=fig2a.eps,width=6.5cm,angle=0}}\parbox{7.cm}{\epsfig{file=fig2b.eps,width=6.5cm,angle=0}}}
\vskip.5em
\centerline{\parbox{7.cm}{\epsfig{file=fig2c.eps,width=6.5cm,angle=0}}}
\caption{\small 
  Probability density of membrane voltage for different orders of magnitude of
  the correlation times $\tau_e, \tau_i$. Panel (b) corresponds to the
  parameter set used by R\&D in Fig.~2b of their paper. Starting from this
  case, we show simulation results with both time constants reduced (panel a)
  or enlarged (panel c) by one order of magnitude.  We compare the simulations
  in all cases to the curve according to the expression given by R\&D (here
  \e{rho_rud}) (dashed lines) and in the limit cases of short and large
  correlation times to the white noise (\e{rho_wn}) and the static noise
  (\e{rho_static}) theories (both solid lines), respectively. Note that the
  expressions for the limit cases fit the simulation data not perfectly but
  much better than the original solution \e{rho_rud} by R\&D. Parameters:
  $g_L=0.0452$mS/cm$^2$, $a=34636 \mu$m$^2$,
  $C_m=1\mu$F/cm$^2$,$E_L=-80$mV,$E_e=0$mV,$E_i=-75$mV,$\sigma_e=0.012\mu$S,
  $\sigma_i=0.0264\mu$S, $g_{e0}=0.0121\mu$S, $g_{i0}=0.0573\mu$S;
  additive-noise parameters ($\sigma_I,I_0$) are all zero, correlation times
  of the noises as indicated. Here we used a time step of $\Delta t=0.1$ ms
  and a simulation time of $100$s as in \citep{RudDes03}.
\label{fig:simu_theo2}}
\end{figure}
For this specific parameter set the agreement is indeed relatively good,
although there are differences between the formula and the simulation results
in the location of the maximum as well as at the flanks of the density. These
differences do not vanish by extending the simulation time or decreasing the
time step; hence, the curve according to \e{rho_rud} does not seem to be an
exact solution but at best a good approximation.\\
The disagreement becomes significant if the correlation times are changed by
one order of magnitude (\bi{simu_theo2}a and c) (in this case we keep the
variances of the noises constant, as R\&D have done rather than the noise
intensities as in \bi{simu_theo}). The asymptotic formulas for either
vanishing (\bi{simu_theo2}a) or infinite (\bi{simu_theo2}c) correlation times
derived here in this paper do a much better job in these limits. Note that the
large correlation time used in \bi{simu_theo2}c are outside the range
considered by R\&D (2003). Regardless of the fact that the correlation times
we have used in \bi{simu_theo2}a and c are possibly outside the physiological
range, an analytical solution should also cover these cases. Regarding the
question of whether the correlation time is short (close to the white-noise
limit), long (close to the static limit), or intermediate (as it seems to be
the case in the original parameter set of Fig.2b in \citep{RudDes03}), it is
not the absolute value of $\tau_{e,i,I}$ that matters but the product
$\beta \tau_{e,i,I}$. Varying one or more of the parameters $g_L,
g_{e0},g_{i0},a,$ or $C_m$ can push the dynamics in one of the limit cases
without the necessity of changing $\tau_{e,i,I}$.\\
\subsection{Probability density of the membrane voltage - extended expression
by R\&D} 
So far we have not considered the extended expression (R\&D, 2005) with the
effective correlation times. Plotting the simulation data shown in
\bi{simu_theo}a and \bi{simu_theo2} against this new formula, gives
indeed a very good although not perfect agreement (cf. \bi{old_new}a and b).
Note, for instance, in \bi{old_new}a that the height of the peak for
$\tau_e=1$ and the location of the maximum for $\tau_e=0.1$ is slightly
underestimated by the new theory. Since most of the data look similar to
Gaussians, we may also check whether they are described by the ETC theory
(cf. \e{rho_etc}).
\begin{figure}[h!]
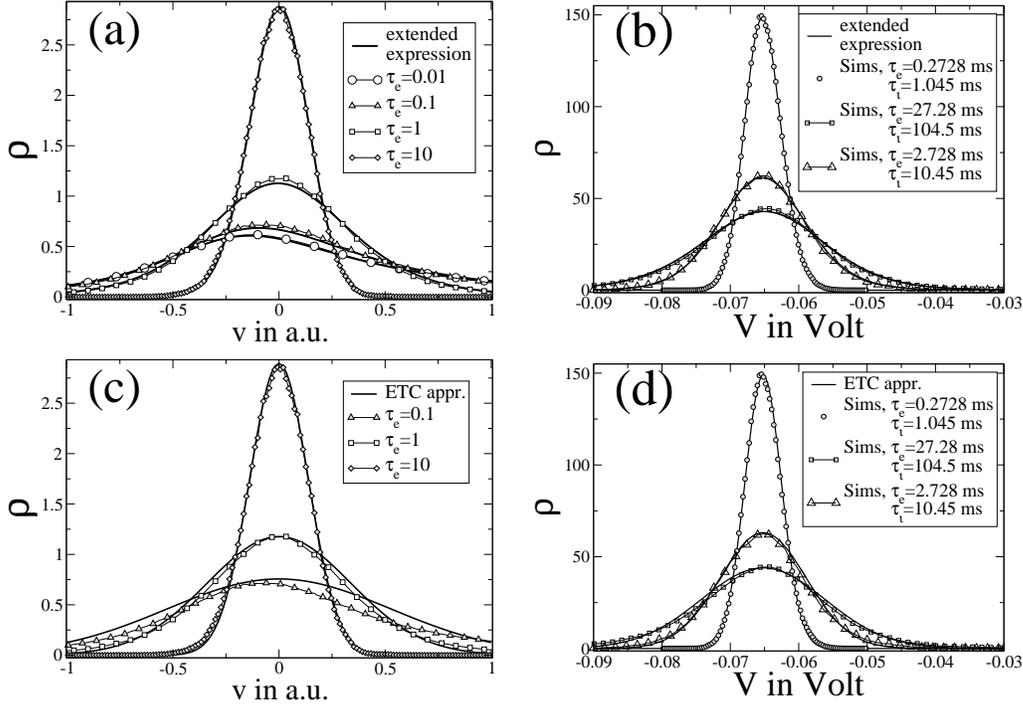

\centerline{\parbox{7.cm}{\epsfig{file=fig3a.eps,width=6.5cm,angle=0}}\parbox{7.cm}{\epsfig{file=fig3b.eps ,width=6.5cm,angle=0}}}
\vskip0.5em
\centerline{\parbox{7.cm}{\epsfig{file=fig3c.eps,width=6.5cm,angle=0}}\parbox{7.cm}{\epsfig{file=fig3d.eps ,width=6.5cm,angle=0}}}
\caption{\small 
  Probability density of membrane voltage for different parameter sets, the
  extended expression \e{rho_rud2} (a and b) and the effective-time constant
  approximation \e{rho_etc} (c and d) are compared to results of numerical
  simulations; simulation data and parameters as in \bi{simu_theo}a (a and c)
  and \bi{simu_theo2} (b and d).
\label{fig:old_new}}
\end{figure}
This is shown in \bi{old_new}c and d and reveals that for the parameter sets
studied so far, the noise intensities are reasonably small such that the ETC
formula gives an approximation almost as good as the extended expression by
R\&D. One exception to this is shown in \bi{old_new}c: at small correlation
times where the noise is effectively white ($\tau_e=0.1$), the ETC formula
fails since the noise variances become large.  For $\tau_e=0.01$ the
disagreement is even worse (not shown). In this range, the extended expression
captures the density better, in particular its non-gaussian features (e.g. the
asymmetry in the density).\\
\begin{figure}[h!]
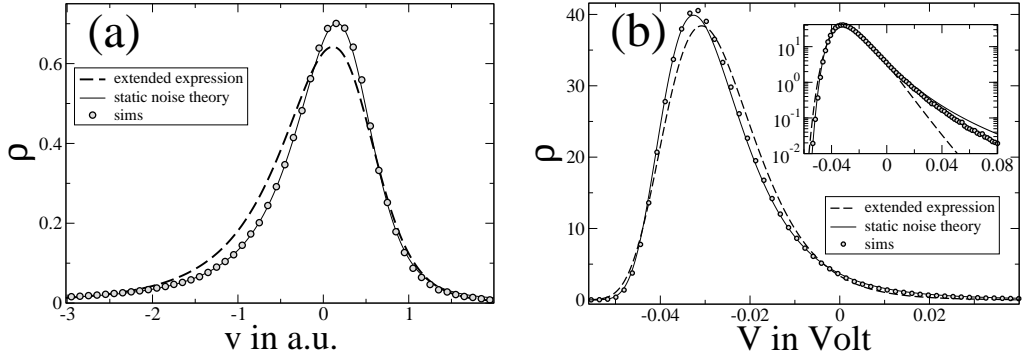

\centerline{\parbox{7.cm}{\epsfig{file=fig4a.eps,width=6.5cm,angle=0}}\parbox{7.cm}{\epsfig{file=fig4b.eps,width=6.5cm,angle=0}}}
\caption{\small 
  Probability density of membrane voltage for long correlation times;
  comparison of static-noise theory (\e{rho_static}, solid lines) and extended
  expression by R\&D (\e{rho_rud2}, dashed lines) to numerical simulations
  (symbols). Panel a: Probability density of the shifted voltage variable (in
  arbitrary units) with $Q_e=Q_i=3, \beta=1,\tau_e=\tau_i=20, V_e=1.5,
  V_i=-0.5$0; for these parameters, the mean value is infinite; in the
  simulation we therefore had to implement reflecting boundaries, which will
  (if sufficiently distant) affect the density only in its tails but not in
  the range shown in the figure. Panel b: density for the original voltage
  variable with $g_L=0.0452$mS/cm$^2$, $a=34636 \mu$m$^2$,
  $C_m=1\mu$F/cm$^2$,$E_L=-80$mV,$E_e=0$mV,$E_i=-75$mV,$\sigma_e=0.012\mu$S,
  $\sigma_i=0.045\mu$S, $g_{e0}=0.121\mu$S, $g_{i0}=0.0574\mu$S,
  $\tau_e=7.5$ms, $\tau_i=30$ms; for these parameters the mean value is
  finite.  Inset: Same data but on a logarithmic scale.
\label{fig:static_limit}}
\end{figure}
Since the agreement of the extended expression to numerical simulations was
so far very good, one could argue that it represents the exact solution to
the problem and the small differences are merely due to numerical
inaccuracy.\\
We will check whether the extended expression is the exact solution in two
ways. Firstly, we know how the density behaves if both multiplicative noises
are very slow ($\beta\tau_e,\beta\tau_i\gg1$), namely, according to
\e{rho_static}. We possess thus an additional control of whether the extended
solution \e{rho_rud2} is exact by comparing not only to numerical simulation
results but also to the static-noise theory.  Secondly, we have derived an
exact integral expression \e{mean_stat_exact} for the stationary mean value;
so we can compare the stationary mean value according to the extended
expression by R\&D (given in \e{mean_stat_rud_ext}) to the exact expression
and to numerical simulations.\\
To check the extended expression against the static-noise theory we have to
choose parameter values for which $\beta\tau_e$ and $\beta\tau_i$ are much
larger than one; at the same time the noise variances should be sufficiently
large. We compare both theories \e{rho_rud2} and \e{rho_static} once for the
system \e{my_v}, \e{my_y} with simplified parameters at strong noise
($Q_e=Q_i=1$) and large correlation times ($\beta\tau_{e,i}=20$)
(\bi{static_limit}a) and once for the original system
(\bi{static_limit}b). For the latter, increases in $\beta\tau_{e,i}$ can be
achieved by either increasing $g_L, g_{e0}, g_{i0}$ or by increasing the
synaptic correlation times $\tau_{e,i}$. We do both and increase $g_{e0}$ to
the ten-fold of the standard value by R\&D (i.e.  $g_{e0}=0.0121\mu$S $\to$
$g_{e0}=0.121\mu$S) and also multiply the standard values of the correlation
times by roughly three (i.e.  $\tau_e=2.728$ms, $\tau_i=10.45$ms $\to$
$\tau_e=7.5$ms, $\tau_i=30$ms); additionally, we also choose a larger standard
deviation for the inhibitory conductance than in R\&D's standard parameter set
($\sigma_i=0.0264\mu$S $\to$ $\sigma_i=0.045\mu$S). For these parameters we
have $\beta\tau_{e}\approx 4.2$ and $\beta\tau_{i}\approx 16.8$, so we may
expect a reasonable agreement between static-noise theory and the true
probability density of the voltage obtained by simulation.\\
Indeed, for both parameter sets, the static-noise theory works reasonably
well.  For the simulation of the original system (\bi{static_limit}b), we
also checked that the agreement is significantly enhanced (agreement within
line width) by using larger correlation times (e.g. $\tau_e=20$ms,
$\tau_i=100$ms) as it can be expected. Compared to the static-noise theory,
the extended expression by R\&D shows stronger although not large
deviations. There are differences in the location and height of the maximum of
the densities for both parameter sets; prominent is also the difference
between the tails of the densities (\bi{static_limit}b inset). Hence, there
are parameters that are not completely outside the physiological range, for
which the extended expression does not yield an exact but only an approximate
description and for which the static-noise theory works better than the
extended expression by R\&D. This is in particular the case for strong and
long-correlated noise.\\
\subsection{Mean value of the membrane voltage}
The second way to check the expressions by R\&D was to compare their mean
values to the exact expression for the stationary mean
\e{mean_stat_exact}.  We do this for the transformed system
\e{my_v}, \e{my_y} with dimensionless parameters. In \bi{mean_vs_tauE},
the stationary mean value is shown as a function of the correlation time
$\tau_e$ of the excitatory conductance.  In the two different panels we keep
the noise intensities $Q_e$ and $Q_i$ fixed; the correlation time of
inhibition is small (panel a) or medium (panel b) compared to the intrinsic
time scale ($1/\beta=1$). We choose noise intensities $Q_i=0.3$ and $Q_e=0.2$
so that the mean value is finite because \e{condi_colored} is satisfied. In
\bi{mean_vs_tauE}a the disagreement between the extended theory by R\&D
(dash-dotted line) and the exact solution (thick solid line) is apparent for
medium values of the correlation time. To verify this additionally, we also
compare to numerical simulation results. The latter agree with our exact
theory for the mean value within the numerical error of the simulation. We
also plot two limits that may help to understand why the new theory by R\&D
works in this special case at very small and very large values of $\tau_E$. At
small values, both noises are effectively white and we have already discussed
that in this case the extended expression for the probability density
\e{rho_rud2} approaches the correct white-noise limit. Hence, also the moment
should be correctly reproduced in this limit. On the other hand, going to
large correlation time $\tau_e$ at fixed noise intensity $Q_e$ means, that the
effect of the colored noise $y_e(t)$ on the dynamics vanishes.  Hence, in this
limit we obtain the mean value of a system that is driven only by one white
noise (i.e. $y_i(t)$). Also this limit is correctly described by R\&D's new
theory since the effective noise intensity $Q'_e=2Q_e/[1+\beta\tau_e]$
vanishes for $\tau_e\to\infty$ if $Q_e$ is fixed. However, for medium values
of $\tau_e$, the new theory predicts a larger mean value than the true
value. The mean value \e{mean_stat_rud} of the original solution \e{rho_rud}
(dotted lines in \bi{mean_vs_tauE}) leads to a mean value of the voltage that
does not depend on the correlation time $\tau_e$ at all.\\
\begin{figure}[h!]
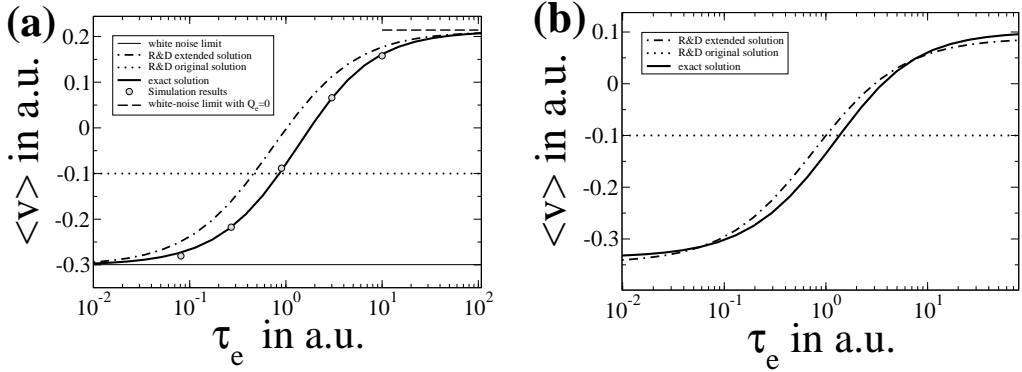

\centerline{\parbox{7.cm}{\epsfig{file=fig5a.eps,width=6.5cm,angle=0}}\parbox{7.cm}{\epsfig{file=fig5b.eps,width=6.5cm,angle=0}}}
\caption{\small 
  Stationary mean value of the shifted voltage (in arbitrary units) vs
  correlation time (in arbitrary units) of the excitatory conductance.  Noise
  intensities $Q_e=0.2,Q_i=0.3, Q_I=0$ and $\beta=1$ are fixed in all
  panels. The correlation time of the inhibitory conductance is
  $\tau_i=10^{-2}$ (a) and $\tau_i=1$ (b). Shown are the exact analytical
  result \e{mean_stat_exact} (solid line), the mean value according to the
  original solution \e{mean_stat_rud} (dotted line), and the mean value
  according to the extended solution \e{mean_stat_rud_ext} (dash-dotted
  line). In panel a, we also compare to the mean value of the white-noise
  solution for $Q_e=0.2,Q_i=0.3$ (thin solid line) and for $Q_e=0,Q_i=0.3$
  (dashed line) as well as to numerical simulation results (symbols).
\label{fig:mean_vs_tauE}}
\end{figure}
If the second correlation time $\tau_I$ is of the order of the effective
membrane time constant $1/\beta$ (\bi{mean_vs_tauE}b), the deviations between
the mean value of the extended expression and the exact solution is smaller
but extends over all values of $\tau_e$. In this case the new solution does
not approach the correct one in neither of the limit cases $\tau_e\to0$ or
$\tau_e\to\infty$.  The overall deviations between the mean according to the
extended expression is small. Also for both panels, the differences in the
mean are small compared to the standard deviations of the voltage. Thus, the
expression \e{mean_stat_rud_ext} corresponding to the extended solution can be
regarded as a good approximation for the mean value.\\
\begin{figure}[h!]
\centerline{\parbox{7.cm}{\epsfig{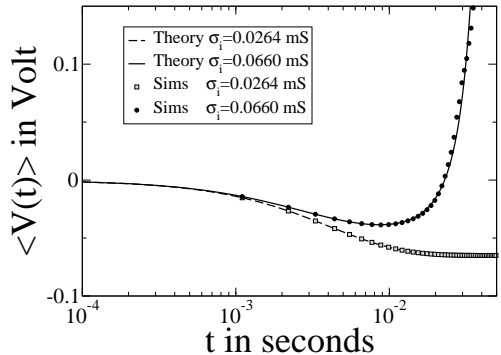}}}
\caption{\small 
Time dependent mean value of the original voltage variable (in Volts) as a
function of time (in seconds) for the initial value $V(t=0)=0$V and different
values of the inhibitory conductance standard deviation $\sigma_i$; numerical
simulations of \e{my_v} and \e{my_y} (circles) and theory according to
\e{mean_direct} (solid lines).  For all curves $g_{e0}=0.0121$ $\mu$S,
$g_{i0}=0.0573$ $\mu$S, $\sigma_{e}=0.012$ $\mu$S, $\tau_{e}=2.728$ ms,
$\tau_{i}=10.49$ ms, $a=34636\mu \mbox{m}^2$, and $C_m=1 \mu$F/cm$^2$. For the
dashed line (theory) and the grey squares (simulations) we choose
$\sigma_{i}=0.0264$ $\mu$S, hence in this case, parameters correspond to the
standard parameter set by \citet{RudDes03}.  For the solid line (theory) and
the black circles we used $\sigma_{i}=0.066$ $\mu$S corresponding to the 250\%
of the standard value by R\&D. While at the standard parameter set, the mean
value saturates at a finite level, in the second case the mean diverges and
goes beyond 100mV within 31ms. Simulations were carried out for $10^6$ voltage
trajectories using an adaptive time step (always smaller than 0.01 ms) that
properly took into account those trajectories that diverge strongest. The
large number of trajectories was required in order to get a reliable estimate
of the time-dependent mean value in the case of strong noise
($\sigma_{i}=0.066$ $\mu$S) where voltage fluctuations are quite large.
\label{fig:mean_vs_t}}
\end{figure}
Finally, we illustrate the convergence or divergence of the mean if the
condition \e{condi_colored} is obeyed or violated, respectively.  First, we
choose the original system and the standard set of parameters by
\citet{RudDes03} and simulate a large number of trajectories in parallel.  All
of these are started at the same value ($V=0$) and each with independent noise
sources, the initial values of which are drawn from the stationary Gaussian
densities. In an experiment, this corresponds exactly to fixing the voltage of
the neuron via voltage clamp and then to let the voltage freely evolve under
the influence of synaptic input (that has not been affected by the voltage
clamp). We compare the time-dependent average of all trajectories to our
theory \e{mean_direct} (in terms of the original variable and parameters). For
R\&D's standard parameters the mean value reaches after a relaxation of
roughly 20ms a finite value (V$\approx-65$mV). The time course of the mean
value is well reproduced by our theory as it should be. Increasing one of the
noise standard deviations to a 2.5-fold of its standard value
($\sigma_{i}=0.0264$ $\mu$S $\to$ $0.066$ $\mu$S), which is still in the range
inspected by R\&D, results in a diverging mean\footnote{These parameter values
were not considered by R\&D to be in the physiological range. We cannot,
however, exclude that other parameter variations (e.g. decreasing the leak
conductance or increasing the synaptic correlation times) will not lead to a
diverging mean for parameters in the physiological range.}. Again the theory
(solid line) is confirmed by the simulation results (black circles). Starting
from zero voltage, the voltage goes beyond 100mV within 31ms. In contrast to
this, the mean value of the extended formula is finite (the condition
\e{condi_rud2} is obeyed) and the mean value formula for this density
\e{mean_stat_rud_ext} yields a stationary mean voltage of $-66$ mV. Thus, in
the general colored-noise case, the extended formula cannot be used to decide
whether the moments of the membrane voltage will be finite or not.\\
We note that the divergence of the mean is due to a small number of strongly
deviating voltage trajectories in the ensemble over which we average. This
implies that the divergence will not be seen in a typical trajectory and that
a large ensemble of realizations and a careful simulation of the rare strong
deviations (adaptive time step) is required to confirm the diverging mean
predicted by the theory. Thus, although the linear model with multiplicative
Gaussian noise is thought to be a simple system compared to nonlinear spike
generators with Poissonian input noise, its careful numerical simulation may
be much harder than that of the latter type of model.
\section{Conclusions}
\label{sec:conclusions}
We have demonstrated that the formula for the probability density of the
membrane voltage driven by multiplicative and/or additive (conductance and/or
current noise) proposed by R\&D in their original paper is wrong in
general. In particular, it fails in all tractable limit cases (white noise
driving, colored additive noise, and static multiplicative noise). Their
extended expression, however, seems to provide a good approximation to the
probability density of the system for a large range of parameters.\\
In the appendix we show where errors have been made in the derivation of the
Fokker-Planck equation on which both the original and extended solutions are
based. Although there are serious flaws in the derivation, we have seen that
the new formula (obtained by an ad-hoc introduction of effective correlation
times in the original solution) gives a very good reasonable approximation to
the probability density for weak noise.  What could be the reason for this
good agreement?\\
The best though still phenomenological reasoning for the solution
\e{rho_rud2} goes as follows. Firstly, an approximation to the
probability density should work in the solvable white-noise limit
\be 
\label{app_wn}
\lim\limits_{\tau_e,\tau_i\to 0}
\rho_{appr}(v,Q_e,Q_i,\tau_e,\tau_i)=\rho_{wn}(v,Q_e,Q_i) 
\ee
Secondly, we know that at weak multiplicative noise of arbitrary
correlation time the effective-time constant approximation will be
approached
\be
\label{app_etc}
\rho_{appr}(v,Q_e,Q_i,\tau_e,\tau_i)=\rho_{ETC}(v,Q_e,Q_i,\tau_e,\tau_i),\;\; (Q_e,Q_i \;\;\mbox{small}) 
\ee 
The latter density given in \e{rho_etc} can be expressed by the white-noise
density with rescaled noise intensities (note that the variance in the ETC
approximation given in \e{var_etc} has this property); furthermore, it is
close to the density for white multiplicative noise if the noise is weak
\ba
\rho_{ETC}(v,Q_e,Q_i,\tau_e,\tau_i)&=&\rho_{ETC}(v,Q_e/(1+\beta\tau_e),Q_i/(1+\beta\tau_i),0,0),\nz\\
 \;\;\;\;\; &\stackrel{(Q_e,Q_i \;\;\mbox{\footnotesize small})}{\approx} &\rho(v,Q_e/(1+\beta\tau_e),Q_i/(1+\beta\tau_i),0,0)\nz\\  
&= &\rho_{wn}(v,Q_e/(1+\beta\tau_e),Q_i/(1+\beta\tau_i)) 
\ea
Hence, using this equation together with \e{app_etc}, one arrives at
\be
\rho_{appr}(v,Q_e,Q_i,\tau_e,\tau_i)\approx\rho_{wn}(v,Q_e/(1+\beta\tau_e),Q_i/(1+\beta\tau_i))
\ee
This approximation which also obeys \e{app_wn} is the extended expression by
R\&D. It is expected to function in the white-noise and the weak-noise limits
and can be regarded as an interpolation formula between these limits.  We have
seen that for stronger noise and large correlation times (i.e. in a parameter
regime where neither of the above assumptions of weak or uncorrelated noise
hold true) this density and its mean value disagree with numerical simulation
results as well as with our static-noise theory. Regarding the parameter sets
for which we checked the extended solution for the probability density, it is
remarkable that the differences to numerical simulations were not stronger.

Two issues remain. Firstly, we have shown that the linear model with Gaussian
conductance fluctuations can show a diverging mean value.  Certainly, for
higher moments, as for instance, the variance, the restrictions on parameters
will be even more severe than that for the mean value (this can be concluded
from the tractable limit cases we have considered). As demonstrated in case of
the stationary mean value, the parameter regime for such a divergence cannot
be determined using the different solutions proposed by R\&D.\\
Of course, a real neuron {\em can} be driven by a strong synaptic input {\em
without} showing a diverging mean voltage --- the divergence of moments found
above is just due to the limitations of the model. One such limitation is the
diffusion approximation on which the model is based.  Applying this
approximation, the synaptically filtered spike train inputs have been replaced
by Ornstein-Uhlenbeck processes. In the original model with spike train input,
it is well known that the voltage cannot go below the lowest reversal
potential $E_i$ or above the excitatory reversal potential $E_e$ if no current
(additive) noise is present (see e.g. \citet{LanLan87} for the case of
unfiltered Poissonian input). In this case, we do not expect a power law
behavior of the probability density at large values of the voltage. Another
limitation of the model considered by R\&D is that no nonlinear spike
generating mechanism has been included. In particular, the mechanism
responsible for the voltage reset after an action potential would prevent any
power law at strong positive voltage.  Thus, we see that at strong synaptic
input the shot-noise character of the input and nonlinearities in the dynamics
cannot be neglected anymore and even determine whether the mean of the voltage
is finite or not.\\
The second issue concerns the consequences of the diffusion approximation for
the validity of the achieved results. Even if we assume a weak noise such that
all the lower moments like mean and variance will be finite, is there any
effect of the shot-noise character of the synaptic input that is not taken
into account properly by the diffusion approximation?  \citet{RicGer05} have
recently addressed this issue and shown that the shot-noise character will
affect the statistics of the voltage and that its contribution is comparable
to that resulting from the multiplicativity of the noise.  Thus, for a
consistent treatment one should either include both features (as done by
\citet{RicGer05} in the limit of weak synaptic noise) or none (corresponding
to the effective-time-scale approximation, cf. \citet{RicGer05}).\\
Summarizing, we believe that the use of the extended expression by
R\&D is restricted to parameters obeying
\be
\label{rud_ext_valid}
\beta\gg Q_e+Q_i.  
\ee
This restriction is consistent with (1) the diffusion approximation on which
the model is based; (2) a qualitative justification of the extended expression
by R\&D as given above; (3) the finiteness of the stationary mean and
variance.\\
For parameters which do not obey the condition \e{rud_ext_valid}, one
should take into account the shot-noise statistics of the synaptic
drive. Recent perturbation results were given by \citet{RicGer05}
assuming weak noise; we note that the small parameter in this theory
is $(Q_e+Q_i)/\beta$ and therefore exactly equal to the small
parameter in \e{rud_ext_valid}.\\
The most promising result in our paper seems to be the exact solution for the
time-dependent mean value, a statistical measure that can be easily determined
in an experiment and might tell us a lot about the synaptic dynamics and its
parameters. The only weakness of this formula is that it is still based on the
diffusion approximation, i.e. on the assumption of Gaussian conductance
noise. One may, however, overcome this limitation by repeating the calculation
for synaptically filtered shot-noise.
\section{Acknowledgments}
This research was supported by NSERC Canada and a Premiers Research Excellence
Award (PREA) from the Government of Ontario. We also acknowledge an anonymous
reviewer for bringing the unpublished Note by R\&D to our attention.
\begin{appendix}
\section{Errors made in the derivation of the Fokker-Planck equation}
\label{sec:app}
Let us first note that although R\&D use a so-called Ito rule, there
is no difference between the Ito and Stratonovich interpretations of
the colored-noise-driven membrane dynamics. Since the noise processes
possess a finite correlation time, the Ito-Stratonovich dilemma
occurring in systems driven by white multiplicative noise is not an
issue here.\\
To comprehend the errors in the analytical derivation of the
Fokker-Planck equation in R\&D, {\em it suffices to consider the case
  of only additive OU noise}. For clarity we will use our own
notation: the OUP is denoted by $y_I(t)$ and we set $h_I=1$ (the
latter function is used in R\&D for generality). R\&D give a formula
for the differential of an arbitrary function $F(v(t))$ in eq.~(B.9)
\be
\label{diffF_rud0}
dF(v(t))=\partial_vF(v(t)) dv
+\frac{1}{2}\partial_v^2F(v(t)) (dv)^2
\ee
R\&D use the membrane equation in its differential form, which for vanishing
multiplicative noises reads
\be
\label{mem_diff}
dv=f(v) dt +dw_I
\ee
where the drift term is $f(v)=-\beta v$ and $w_I$ is the integrated OU process
$y_I$, i.e.
\be
w_I=\intl_0^t ds\; y_I(s).
\ee
Inserting \e{mem_diff} into \e{diffF_rud0}, we obtain
\be
\label{diffF_rud}
dF(v(t))=\partial_vF(v(t)) f(v(t)) dt +\partial_v F(v(t)) d\tilde{w}_I
+\frac{1}{2}\partial_v^2F(v(t)) (dw_I)^2
\ee
This should correspond to eq.~(B.10) in R\&D for the case of zero
multiplicative noise. However, our formula differs from eq.~(B.10) in one
important respect: R\&D have replaced $(dw_I)^2$ by $2\alpha_I(t) dt$ using
their Ito rule\footnote{Note that R\&D use $\alpha_I(t)$ for two different
expressions, namely, according to eq.~(B.8) for
$\tilde{\sigma}_I^2[\tau_I(1-\exp(-t/\tau_I))-t]+w_I^2(t)/(2\tau_I)$ but also
according to eq.~(3.2) in R\&D for the average of this stochastic quantity.} 
eq.~(A.13a).  Dividing by $dt$, averaging, and using the fact that for finite
$\tau_I$ $dw_I(t)/dt=y_I(t)$, we arrive at
\be
\label{dglF_rud}
\frac{d\lr{F(v(t))}}{dt}=\lr{\partial_v F(v(t)) f(v(t))} +\lr{\partial_v
  F(v(t)) y_I(t)} +\frac{1}{2}\LR{\partial_v^2F(v(t)) \frac{(dw_I)^2}{dt}}.
\ee
This should correspond to eq.~(B.12) in R\&D (again for the case of
vanishing multiplicative noise) but is not equivalent to the latter
equation for two reasons.  Firstly, R\&D set the second term on the
r.h.s. to zero reasoning that the mean value $\lr{y_I(t)}$ is zero
(they also use an argument about $h_{\{e,i,I\}}$, which is irrelevant
in the additive noise case considered here).  Evidently, if $y_I(t)$
is a colored noise it will be correlated to its values in the past
$y(t')$ with $t'<t$.  The voltage $v(t)$ and any nontrivial function
$F(v(t))$ is a functional of and therefore correlated to $y_I(t')$
with $t'<t$.  Consequently, there is also a correlation between
$y_I(t)$ and $F(v(t))$, and thus
\be
\label{correlation}
\lr{\partial_v F(v(t)) y_I(t)}\neq\lr{\partial_v F(v(t))}\lr{y_I(t)}=0
\ee
Hence, setting the second term (that actually describes the effect of the
noise on the system) to zero is wrong\footnote{For those readers, still
unconvinced of \e{correlation} a simple example: Let $F(v(t))=v^2(t)/2$. Then
$\lr{\partial_v F(v(t)) y_I(t)}=\lr{v(t) y_I(t)}$. In the stationary state
this average can be calculated as $\int\int dv dy_I\; v y_I P_0(v,y_I)$ using
the density \e{rho_vy}. This yields $\lr{v(t) y_I(t)}=Q_I/[1+\beta\tau_I]$
which is finite for all finite values of the noise intensity $Q_I$ and
correlation time $\tau_I$. Note that this line of reasoning is only valid for
truly colored noise ($\tau_I>0$); the white-noise case has to be treated
separately.}.  This also applies for the respective terms due to the
multiplicative noise.\\
Secondly, the last term on the r.h.s. of \e{dglF_rud} was treated as a
finite term in the limit $t\to\infty$. According to R\&D's eq.~(A.13a)
(for $i=j$), eq.~(3.2), and eq.~(3.3), $\lim\limits_{t\to\infty}
\lr{(dw_I)^2}= \lim\limits_{t\to\infty} 2\alpha_I(t) dt=\tilde{\sigma}_I^2\tau_I
dt$ and, thus $\lr{(dw_I^2)}/dt\to \tilde{\sigma}_I^2\tau_I$ as
$t\to\infty$.  However, the averaged variance of $dw_I=y_I(t)dt$ is
$\lr{(dw_I)^2}=\lr{y_I(t)^2} (dt)^2=\tilde{\sigma}_I^2 (dt)^2$ and,
therefore, the last term in \e{dglF_rud} is of first order in $dt$
(since $(dw_I)^2/dt=y_I(t)^2 dt \sim dt$) and vanishes. We
note that the limit in eq.~(3.3) is not correctly carried out --- even
if we follow R\&D in using their relations (A.13a) together with the
correct relation (A.10a), we obtain that for finite $\tau_I$, the mean
squared increment $\lr{(dw_I)^2}$ is zero in linear order in $dt$ for
all times $t$ which is in contradiction to eq.~(3.3) in R\&D.\\
We now show that keeping the proper terms in \e{dglF_rud} does not
lead to a useful equation for the solution of the original problem.
After applying what was explained above, \e{dglF_rud} reads correctly
\be
\label{dglF}
\frac{d\lr{F(v(t))}}{dt}=\lr{\partial_v F(v(t)) f(v(t))} +\lr{\partial_v
  F(v(t)) y_I(t)}. 
\ee
Because of the correlation between $v(t)$ and $y_I(t)$, we have to use
the full two-dimensional probability density to express the averages
\ba
\label{av_rels}
\lr{\partial_v F(v(t)) f(v(t))}&=&\int dv \int dy_I (\partial_v F(v)) f(v) P(v,y_I,t)\nz\\
&=&\int dv (\partial_v F(v)) f(v) \rho(v,t) \nz\\
\lr{\partial_v F(v(t))y_I(t)}&=& \int dv \int dy_I (\partial_v F(v))
y_I P(v,y_I,t) 
\ea
Inserting these relations into \e{dglF}, performing an integration by
part, and setting $F(v)=1$ leads us to
\be
\partial_t \rho(v,t)=-\partial_v (f(v) \rho(v,t)) -\partial_v \left(\int dy_I y_I P(v,y,t)\right) 
\ee
which is not a closed equation for $\rho(v,t)$, nor is it a
Fokker-Planck equation. The above equation with $f(v)=-\beta v$ can be
also obtained by integrating the two-dimensional Fokker-Planck
equation \e{P_vy} over $y_I$.\\
In conclusion, by neglecting a finite term and assuming a vanishing term to be
finite, R\&D have effectively {\em replaced} one term by the other, i.e.  the
colored-noise drift term is replaced by a white-noise diffusion term, the
latter with a prefactor that corresponds to only half of the noise
intensity. This amounts to a white-noise approximation of the colored
conductance noise, although with a noise-intensity that is not correct in the
white-noise limit of the problem. 
\end{appendix}
\bibliographystyle{plainnat} \bibliography{../../../BIB/ALL}
\end{document}